\pdfoutput=1

\documentclass[12pt,usenames,dvipsnames]{article}

\usepackage{latexsym}
\usepackage{amssymb, amsfonts, amsmath}
\usepackage{graphicx}
\usepackage{indentfirst}
\usepackage{bbm}
\usepackage{amssymb}
\usepackage{verbatim}
\usepackage{amsmath, amsthm, amssymb}
\usepackage{mathrsfs}
\usepackage{hyperref}
\usepackage{amsfonts}
\usepackage{dsfont}
\usepackage{cite}
\usepackage{xcolor}
\usepackage{enumerate}
\usepackage{cleveref}
\usepackage{physics}
\usepackage{cases}

\parskip=\medskipamount

\newcommand {\cL}{{\cal L}}

\newcommand {\cN}{{\cal N}}

\def\a{\alpha}

\def\b{\beta}
\def\c{\chi}
\def\d{\delta}

\def\f{\phi}

\def\j{\psi}

\def\l{\lambda}
\def\m{\mu}
\def\n{\nu}
\def\o{\omega}

\def\r{\rho}
\def\s{\sigma}

\def\x{\xi}

\def\rd{{\rm d}}
\def\ri{{\rm i}}
\def\re{{\rm e}}

\newcommand{\pa}{\partial}                           
\newcommand{\hf}{\frac12}

\newcommand{\vf}{\varphi}

\newcommand{\be}{\begin{equation}}
\newcommand{\ee}{\end{equation}}
\newcommand{\bea}{\begin{eqnarray}}
\newcommand{\eea}{\end{eqnarray}}
\newcommand{\non}{\nonumber}
\def\double #1{#1{\hbox{\kern-2pt $#1$}}}

\newif\ifdtup

\newcommand{\bsubeq}{\begin{subequations}}
\newcommand{\esubeq}{\end{subequations}}

\numberwithin{equation}{section}

\begin{document}

\begin{titlepage}
\begin{flushright}
October, 2023 \\
\end{flushright}
\vspace{5mm}

\begin{center}
{\Large \bf 
Quantisation of the gauge-invariant models for massive higher-spin bosonic fields
}
\end{center}

\begin{center}

{\bf John H. Fegebank and Sergei M. Kuzenko} \\
\vspace{5mm}

\footnotesize{
{\it Department of Physics M013, The University of Western Australia\\
35 Stirling Highway, Perth W.A. 6009, Australia}}  
~\\
\vspace{2mm}
~\\
Email: \texttt{ 
22705104@student.uwa.edu.au, sergei.kuzenko@uwa.edu.au}\\
\vspace{2mm}

\end{center}

\begin{abstract}
\baselineskip=14pt
In 2001, Zinoviev proposed a gauge-invariant formulation for a massive bosonic field with spin $s \geq 2$ in a spacetime  of constant curvature. In this paper we carry out the Faddeev-Popov quantisation of this theory in $d$-dimensional Minkowski space. 
We also make use of the Zinoviev theory to derive a generalisation of the Singh-Hagen
model for a massive integer-spin field in $d>4 $ dimensions.
\end{abstract}
\vspace{5mm}

\vfill

\vfill
\end{titlepage}

\newpage
\renewcommand{\thefootnote}{\arabic{footnote}}
\setcounter{footnote}{0}

\tableofcontents{}
\vspace{1cm}
\bigskip\hrule

\allowdisplaybreaks


\section{Introduction}
\label{Intro}

Long ago \cite{Dirac,Fierz,FierzPauli}, the equations describing an on-shell massive field of arbitrary spin $s$ in four dimensions were derived. In the integer-spin case, the equations are 
\begin{subequations}
\bea
\pa^\n \vf_{\n \m_1 \dots \m_{s-1}} =0~, \qquad (\Box -m^2 ) \vf_{\m_1 \dots \m_s} =0~, \label{Fierz Pauli}
\eea 
where the dynamical field $\vf_{\m_1 \dots \m_s} $ is symmetric and traceless, 
\bea
 \vf_{(\m_1 \dots \m_s)}= \vf_{\m_1 \dots \m_s} 
~, \qquad \eta^{\n \r} \vf_{\n \r \m_1 \dots \m_{s-2}} =0~.
\eea 
\end{subequations}
In order to realise these equations as Euler-Lagrange equations in a Lagrangian field theory, it was pointed out 
that certain auxiliary fields are required for $s>1$ \cite{FierzPauli}.
The correct set of auxiliary fields and the action principle were found by Singh and Hagen in the bosonic \cite {Singh:1974}
and the fermionic \cite{SinghHagen2} cases. One may think of the integer-spin model of \cite {Singh:1974} as a higher-spin generalisation of  the massive spin-two model proposed by Fierz and Pauli
\cite{FierzPauli}, 
\bea
\mathcal{L} &= &\frac{1}{2}\varphi^{\mu \n}(\square-m^2)\varphi_{\mu \n}+\partial_\nu\varphi^{\nu\mu}\partial^\lambda\varphi_{\lambda\mu} 
   +\frac{1}{3}\vf \Big\{ 2 \partial_\mu\partial_\nu\varphi^{\mu\nu}
   - \big(\square-2 m^2\big)\varphi \Big\}~.
\eea
Considering a massless limit of the  Singh-Hagen model allowed Fronsdal to derive gauge-invariant formulations 
for massless higher-spin bosonic fields
\cite{Fronsdal:1978}.

It is well known that the massive spin-one model (known as the Proca theory)
\bea
\cL = - \frac 14 F^{\m \n} F_{\m\n } - \hf m^2 A^\m A_\m ~, \qquad F_{\m\n} = \pa_\m A_\n - \pa_\n A_\m ~,
\label{1.3}
\eea
has a gauge-invariant Stueckelberg  reformulation
\bea
\widetilde{\cL} = - \frac 14 F^{\m \n} F_{\m\n } - \hf \pa^\m \vf \pa_\m \vf - \hf m^2 A^\m A_\m +m A^\m \pa_\m \vf~,
\label{1.4}
\eea
which is obtained from \eqref{1.3} by replacing $A_\m \to A_\m - m^{-1} \pa_\m \vf$. By construction, the gauge symmetry of $\widetilde{\cL}$ has the form 
\bea
\d A_\m = \pa_\m \x~, \qquad \d \vf = m \x~,
\eea
with the gauge parameter $\x $ being arbitrary. This local symmetry allows one to choose the gauge condition $\vf =0$, and then $\widetilde{\cL}$ turns into \eqref{1.3}. 

The Singh-Hagen model \cite {Singh:1974} is a non-gauge theory. For various reasons, it is of interest to have its gauge-invariant reformulation being similar to that for the massive spin-one model just discussed. For the spin values $s=2$ and $s=3$, such reformulations were derived forty years ago by Zinoviev \cite{Zinovev1983}. In 1997, his results were extended by Klishevich and Zinoviev  to the case of an arbitrary integer spin in four-dimensional Minkowski space in \cite{Klishevich:1997pd}.\footnote{It was also shown by the authors of \cite{Klishevich:1997pd} that the Singh-Hagen theory is obtained from their formulation by appropriately fixing the gauge freedom.}
Finally, a gauge-invariant formulation for massive particles of arbitrary integer spin 
was constructed by Zinoviev \cite{Zinoviev:2001}
in a $d$-dimensional space of constant curvature. The construction of \cite{Zinoviev:2001} inspired Metsaev to propose a gauge-invariant formulation for massive totally symmetric fermionic fields in $d$-dimensional (anti-)de Sitter space
\cite{Metsaev}. The gauge-invariant formulations of \cite{Zinoviev:2001,Metsaev}
  have been used for various applications and generalisations, including the frame-like gauge invariant formulation for massive high spin fields \cite{Zinoviev:2008ze} and Lagrangian descriptions of massive $\cN=1$ supermultiplets with arbitrary superspin \cite{Zinoviev:2007js, Buchbinder:2019dof}.

Recently, gauge-invariant actions for massive arbitrary spin particles in $d$ dimensions have been derived \cite{Lindwasser} using  dimensional reduction of the massless Fronsdal's models in $(d+1)$ dimensions
\cite{Fronsdal:1978,FF}. A precise correspondence with the formulations developed earlier in \cite{Zinoviev:2001,Metsaev} has not been discussed.

 To the best of our knowledge, covariant quantisation of the Zinoviev theory has never been studied. This paper is aimed at filling the gap by carrying out 
the Faddeev-Popov quantisation of the theory in Minkowski space ${\mathbb M}^d$. 

This paper is organised as follows. In Section \ref{Zinoviev Action} we review 
  the Zinoviev theory in ${\mathbb M}^d$. Section \ref{Singh Hagen Correspondence} derives a $d$-dimensional extension of the Singh-Hagen theory.\footnote{It was recently mentioned \cite{Lindwasser}
  that ``this formulation only works in four spacetime dimensions.'' } 
 Sections \ref{Spin 2} and \ref{Spin 3} are devoted to the quantisation of the massive spin-2 and spin-3 models.
  The quantisation procedure is then extended  in Section \ref{Spin s}
  to the spin-$s$ case.

Throughout this paper, 
the mostly plus Minkowski metric is used, $\eta_{\m\n}=\text{diag}(-1,+1,\dots,+1)$. 
 The dynamical variables of the spin-$s$ Zinoviev's theory are symmetric double traceless gauge fields 
 $\f^{(k)}_{\m_1 \dots \m_k} $, with $k = s, s-1, \dots , 0$. 
 Given such a field $\phi^{(k)}_{\m_1 \dots \m_k}$, with $k\geq 2$, its trace is denoted
\be
    \Tilde{\phi}^{(k)}_{\m_3 \dots \m_k} := \eta^{\r\s} \f^{(k)}_{\r\s \m_3\dots \m_k}~.
   \ee
 In the $k>0$ case, associated with  $\f^{(k)}_{\m_1 \dots \m_k} $
 is the symmetric and traceless gauge parameter $\x^{(k-1)}_{\m_1 \dots \m_{k-1}}$.
 Let $\overline{\psi}^{ (k-1 )} $ and $\psi^{(k-1)}$ be the symmetric and traceless Faddeev-Popov ghosts
corresponding to  $\x^{(k-1)}$.
  For path integrals, 
 the following shorthand is adopted
\bsubeq
\begin{align}
    \int\mathcal{D}(\phi;k) &:= \int\prod^{k}_{j=0}\mathcal{D}\phi^{(j)}~,\\
    \noalign{\vspace{10pt}}\int\mathcal{D}(\phi,\psi;k) &:= \int\Big(\prod^{k}_{j=1}\mathcal{D}\phi^{(j)}\mathcal{D}\overline{\psi}^{ (j-1 )}\mathcal{D}\psi^{(j-1)}\Big)\mathcal{D}\phi^{(0)}~.
\end{align}
\esubeq
In order to evaluate the ghost contributions, the following identity is noted:
Faddeev-Popov determinants are realised in terms of the path integral according to the general rule:
\begin{subequations} \label{Determinant of M} 
\begin{align}
&       \det M^{(k)}
 = \int \mathcal{D}\overline{\psi}^{(k)}\mathcal{D}\psi^{(k)}\non \\
       &\qquad \qquad  \times \exp\left\{-\ri \int \rd^dx \int \rd^dx'\,\overline{\psi}^{(k)\mu_1\dots\mu_k}(x)
       M_{\mu_1\dots \mu_{k}}{}^{\n_1 \dots \n_k} (x,x') \,\psi^{(k)}_{\nu_{1}\dots\nu_k}(x')\right\}~. 
\end{align}
Here  $M^{(k)}$ is an operator acting on the space of symmetric and traceless fields $\j^{(k)}$, 
\bea
M^{(k)}: \psi^{(k)}_{\mu_{1}\dots\mu_k}(x) \to
\int \rd^dx'\, M_{\mu_1\dots \mu_{k}}{}^{\n_1 \dots \n_k} (x,x')\, \psi^{(k)}_{\nu_{1}\dots\nu_k}(x')~.
\eea
\end{subequations}


\section{The Zinoviev theory} \label{Zinoviev Action}

In this section we briefly review the Zinoviev theory for a massive spin-$s$ field in ${\mathbb M}^d$. It  is described by the action 
\begin{subequations}  \label{Full Action for spin s}
\begin{equation}
    S = \int \rd^dx\,\mathcal{L}^{(s)}~,\qquad 
      \mathcal{L}^{(s)} = \sum^{s}_{k=0}\mathcal{L}_c(\phi^{(k)})~,
 \end{equation}
where the dynamical variables $\f^{(k)}_{\m_1 \dots \m_k} $ are symmetric double traceless fields, and 
$  \mathcal{L}_c(\phi^{(k)}) $ has the form
\begin{equation}
    \mathcal{L}_c(\phi^{(k)}) = \mathcal{L}_0(\phi^{(k)})+\mathcal{L}_m(\phi^{(k)})~.\label{Combined Lagrangian Contribution}
\end{equation}
\end{subequations}
Here the first term on the right is Fronsdal's Lagrangian \cite{Fronsdal:1978} for a massless spin-$k$ field,
\begin{equation}
    \begin{split}
        \mathcal{L}_0(\phi^{(k)})  = & -\frac{1}{2}\partial^\mu\phi^{(k)\mu_1\dots\mu_k}\partial_\mu\phi^{(k)}_{\mu_1\dots\mu_k}+\frac{k}{2}\partial_\mu\phi^{(k)\mu\mu_2\dots\mu_k}\partial^\nu\phi^{(k)}_{\nu\mu_2\dots\mu_k}\\
        & +\frac{k(k-1)}{4}\partial^\mu\Tilde{\phi}^{(k)\mu_3\dots\mu_k}\partial_\mu\Tilde{\phi}^{(k)}_{\mu_3\dots\mu_k}+\frac{k(k-1)}{2}\partial_\mu\partial_\nu\phi^{(k)\mu\nu\mu_3\dots\mu_k}\Tilde{\phi}^{(k)}_{\mu_3\dots\mu_k}\\
        & +\frac{k(k-1)(k-2)}{8}\partial_\mu\Tilde{\phi}^{(k)\mu\mu_4\dots\mu_k}\partial^\nu\Tilde{\phi}^{(k)}_{\nu\mu_4\dots\mu_k}~.
    \end{split}\label{Massless Lagrangian Contribution}
\end{equation}
The second term in \eqref{Combined Lagrangian Contribution} is 
a massive contribution of the following structure:
\begin{equation}
    \begin{split}
        \mathcal{L}_m(\phi^{(k)})  = & a_k\phi^{(k-1)\mu_2\dots\mu_{k}}\partial^\mu\phi^{(k)}_{\mu\mu_2\dots\mu_k}+b_k\Tilde{\phi}^{(k)\mu_3\dots\mu_k}\partial^\mu\phi^{(k-1)}_{\mu\mu_3\dots\mu_k}\\
        & +c_k\partial_\mu\Tilde{\phi}^{(k)\mu\mu_4\dots\mu_k}\Tilde{\phi}^{(k-1)}_{\mu_4\dots\mu_k}+d_k\phi^{(k)\mu_1\dots\mu_k}\phi^{(k)}_{\mu_1\dots\mu_k}\\
        & +e_k\Tilde{\phi}^{(k)\mu_3\dots\mu_k}\Tilde{\phi}^{(k)}_{\mu_3\dots\mu_k}-f_k\Tilde{\phi}^{(k)\mu_3\dots\mu_k}\phi^{(k-2)}_{\mu_3\dots\mu_k}~.
    \end{split}\label{k^th Massive Lagrangian Contribution}
\end{equation}
It involves several numerical coefficients, all of which are determined by requiring the action \eqref{Full Action for spin s} to be invariant under gauge transformations 
\begin{equation}
    \begin{split}
        \delta\phi^{(k)}_{\mu_1\dots\mu_k} =~ & \partial_{(\mu_1}\xi^{(k-1)}_{\mu_2\dots\mu_k)}+\alpha_k\xi^{(k)}_{\mu_1\dots\mu_k}-\frac{k(k-1)}{2}\beta_k\eta_{(\mu_1\mu_2}m\xi^{(k-2)}_{\mu_3\dots\mu_k)} ~,
    \end{split}\label{k^th Gauge Transformation}
\end{equation}
where $\a_k$ and $\b_k$ are some coefficients,
and $\xi^{(k)}$ are symmetric and traceless gauge parameters.\footnote{As usual, it is assumed 
in \eqref{k^th Gauge Transformation} that when a term would contain fields of negative rank in the above, that term is ignored. }
Direct calculations show that  the coefficients in \eqref{k^th Massive Lagrangian Contribution} 
and $\b$'s in \eqref{k^th Gauge Transformation} are fixed in terms of $\a$'s:
\begin{subequations}\label{Zinoviev Coefficients}
\bea
        a_k &=& -k\alpha_{k-1}~,\hspace{20pt} b_k = -k(k-1)\alpha_{k-1}~, \hspace{20pt}c_k = -\frac{k(k-1)(k-2)}{4}\alpha_{k-1}~,\\
             2d_k &=& \frac{2(k+1)(2k+d-3)}{2d+k-4}(\alpha_{k})^2-k(\alpha_{k-1})^2~, 
        \quad        k\geq1~; \qquad  d_0 = \frac{d}{d-2}(\alpha_1)^2~, \\
        2e_k &=& -\frac{k(k^2-1)(2k+d)}{4(2k+d-4)}(\alpha_k)^2+\frac{k^2(k-1)}{2}(\alpha_{k-1})^2~,\\
        2f_k &=& -k(k-1)\alpha_{k-1}\alpha_{k-2}~, \\
        \beta_k &=& \frac{2\alpha_{k-1}}{(k-1)(2k+d-6)}~.
\eea
\end{subequations}
For the coefficients $\a$'s one gets
\begin{subequations} \label{2.6}
\bea
(\alpha_k)^2 = \frac{s(s-k)(s+k+d-3)}{(k+1)(2k+d-2)}(\a_{s-1})^2~, \hspace{20pt} 0\leq k\leq s-2~,
\eea
for which it is convenient to choose $(\a_{s-1})^2$:
\bea
\label{Alpha k}
    (\alpha_{s-1})^2 = \frac{m^2}{s}\quad \implies \quad
    (\alpha_k)^2 = \frac{(s-k)(s+k+d-3)}{(k+1)(2k+d-2)}m^2~, \hspace{20pt} 0\leq k\leq s-2~.
\eea
\end{subequations}


\section{Singh-Hagen model in $d$ dimensions}
\label{Singh Hagen Correspondence}

In four dimensions, it was shown by Klishevich and Zinoviev \cite{Klishevich:1997pd} that the Singh-Hagen theory 
 \cite{Singh:1974}  is obtained from the $d=4$ version of the theory described in the previous section by appropriately fixing the gauge freedom.
 Here we extend the analysis of \cite{Klishevich:1997pd} to $d$ dimensions. 

In order to derive a $d$-dimensional analogue of the Singh-Hagen theory from Zinoviev's one, it suffices to choose a unitary gauge. To start with, for $k>1$ we decompose each double traceless symmetric field $\f^{(k)}_{\m_1 \dots \m_k}$ into a sum of two traceless symmetric fields
$\omega^{(k)}_{\mu_1\dots\mu_k}$ and 
${\varphi}^{(k-2)}_{\mu_1\dots\mu_{k-2}}$. In terms of these, the field  $\f^{(k)}$ is
\begin{equation}
    \phi^{(k)}_{\mu_1\dots\mu_k}=\omega^{(k)}_{\mu_1\dots\mu_k}+
       \frac{k(k-1)}{2(d+2k-4)}
    \eta_{(\mu_1\mu_2}{\varphi}^{(k-2)}_{\mu_3\dots\mu_k)}~,
\end{equation}
with  $\tilde{\f}^{(k)}_{\mu_1\dots\mu_{k-2}} = {\varphi}^{(k-2)}_{\mu_1\dots\mu_{k-2}}$, with $k=s, \dots ,2$.
Then, the gauge freedom \eqref{k^th Gauge Transformation} may be completely fixed by imposing the conditions 
\bea
\omega^{(k)}_{\mu_1\dots\mu_k} =0~, \qquad k = 0, 1, \dots , s-1~.
\label{gauge_condition}
\eea
The remaining rank-$s$ field will be re-labelled as follows: $\omega^{(s)}_{\mu_1\dots\mu_s} =\varphi^{(s)}_{\mu_1\dots\mu_s}$.
As a result, upon imposing the gauge condition \eqref{gauge_condition} we stay with the following fields
\begin{subequations} \label{unitary}
\bea
\phi^{(s)}_{\mu_1\dots\mu_s}&=&\varphi^{(s)}_{\mu_1\dots\mu_s}+\frac{s(s-1)}{2(d+2s-4)}\eta_{(\mu_1\mu_2}\varphi^{(s-2)}_{\mu_3\dots\mu_s)}~,\\
  \phi^{(k)}_{\mu_1\dots\mu_k}&=&
     \frac{k(k-1)}{2(d+2k-4)}
  \eta_{(\mu_1\mu_2}\varphi^{(k-2)}_{\mu_3\dots\mu_k)}~, \qquad 
  k\leq s-1~.
\eea
\end{subequations}

We now turn to massaging the separate contributions to the Lagrangian 
$\mathcal{L}^{(s)}$, eq. 
\eqref{Full Action for spin s}, in the gauge \eqref{gauge_condition} 
or, equivalently, \eqref{unitary}.
For the massless Lagrangian $ \mathcal{L}_0(\phi^{(s)}) $ we obtain
\begin{align}
\label{Massless Spin s Lagrangian Varphi}
        \mathcal{L}_0(\phi^{(s)}) 
             = & \frac{1}{2}\varphi^{(s)\mu_1\dots\mu_s}\square\varphi^{(s)}_{\mu_1\dots\mu_s}
        +\frac{s}{2}\partial_\mu\varphi^{(s)\mu\mu_2\dots\mu_s}\partial^\nu\varphi^{(s)}_{\nu\mu_2\dots\mu_s} \non \\
&        +\frac{s(s-1)(d+2s-6)}{2(d+2s-4)}\partial_\mu\partial_\nu\varphi^{(s)\mu\nu\mu_3\dots\mu_s}\varphi^{(s-2)}_{\mu_3\dots\mu_s} \non \\
        & -\frac{s(s-1)(d+2s-6)(d+2s-5)}{4(d+2s-4)^2}\varphi^{(s-2)\mu_3\dots\mu_s}\square\varphi^{(s-2)}_{\mu_3\dots\mu_s}\\
        & +\frac{(s-2)(d+2s-6)(d+2s-8)s(s-1)}{8(d+2s-4)^2}\partial_\mu\varphi^{(s-2)\mu\mu_4\dots\mu_s}\partial^\nu\varphi^{(s-2)}_{\nu\mu_4\dots\mu_s}~.\non
\end{align}
For $0\leq k\leq s-1$, the massless Lagrangian $ \mathcal{L}_0(\phi^{(k)})  $ leads to 
\begin{align}
        \mathcal{L}_0(\phi^{(k)})  = & -\frac{k(k-1)(d+2k-6)(d+2k-5)}{4(d+2k-4)^2}\varphi^{(k-2)\mu_3\dots\mu_k}\square\varphi^{(k-2)}_{\mu_3\dots\mu_k} \non \\
        &+\frac{k(k-1)(k-2)(d+2k-6)(d+2k-8)}{8(d+2k-4)^2}\partial_\mu\varphi^{(k-2)\mu\mu_4\dots\mu_k}\partial^\nu\varphi^{(k-2)}_{\nu\mu_4\dots\mu_k}~.
\end{align}
Next, we turn to the massive contributions 
\eqref{k^th Massive Lagrangian Contribution}.
For $0\leq k \leq s-1$, we obtain
\begin{align}
        \mathcal{L}_m(\phi^{(k)})  = & a_k\frac{(k-1)(k-2)}{2(d+2k-6)}\varphi^{(k-3)\mu_4\dots\mu_k}\partial^\mu\varphi^{(k-2)}_{\mu\mu_4\dots\mu_k}+b_k\frac{k-2}{d+2k-6}\varphi^{(k-2)\mu\mu_4\dots\mu_k}\partial_\mu\varphi^{(k-3)}_{\mu_4\dots\mu_k} \non \\
        &+c_k\partial_\mu\varphi^{(k-2)\mu\mu_4\dots\mu_k}\varphi^{(k-3)}_{\mu_4\dots\mu_k}+d_k\frac{k(k-1)}{2(d+2k-4)}\varphi^{(k-2)\mu_3..\mu_k}\varphi^{(k-2)}_{\mu_3..\mu_k}\\
        & +e_k\varphi^{(k-2)\mu_3\dots\mu_k}\varphi^{(k-2)}_{\mu_3\dots\mu_k}~.
        \non
\end{align}
Substituting in the values for $a_k$, $b_k$ and $c_k$ from (\ref{Zinoviev Coefficients}), 
upon integration by parts we get
\begin{align}
        \mathcal{L}_m(\phi^{(k)})  = & \alpha_{k-1}\frac{k(k-1)(k-2)(d+2k-8)}{4(d+2k-6)}\varphi^{(k-3)\mu_4\dots\mu_k}\partial^\mu\varphi^{(k-2)}_{\mu\mu_4\dots\mu_k}\non \\
        & +\left(d_k\frac{k(k-1)}{2(d+2k-4)}+e_k\right)\varphi^{(k-2)\mu_3..\mu_k}\varphi^{(k-2)}_{\mu_3..\mu_k}~.
 \end{align}
On the other hand, the massive contribution $ \mathcal{L}_m(\phi^{(s)})  $ is
\begin{align}
\label{Massive Spin s Lagrangian Varphi}
           \mathcal{L}_m(\phi^{(s)})  = & d_s\varphi^{(s)\mu_1\dots\mu_s}\varphi^{(s)}_{\mu_1..\mu_s}+\alpha_{s-1}\frac{s(s-1)(s-2)(d+2s-8)}{4(d+2s-6)}\varphi^{(s-3)\mu_4\dots\mu_s}\partial^\mu\varphi^{(s-2)}_{\mu\mu_4\dots\mu_s} \non \\
        & +\left(d_s\frac{s(s-1)}{2(d+2s-4)}+e_s\right)\varphi^{(s-2)\mu_3..\mu_s}\varphi^{(s-2)}_{\mu_3..\mu_s}~.
   \end{align}

Finally, it remains to substitute the contributions
 (\ref{Massless Spin s Lagrangian Varphi}--\ref{Massive Spin s Lagrangian Varphi}) 
in the  full Lagrangian (\ref{Full Action for spin s}) as well as to make use of \eqref{Zinoviev Coefficients}
and \eqref{2.6}
to end up with 
\begin{align}
    \mathcal{L}^{(s)} = &\frac{1}{2}\varphi^{(s)\mu_1\dots\mu_s}(\square-m^2)\varphi^{(s)}_{\mu_{1}\dots\mu_{s}}+\frac{1}{2}s\partial^\nu\varphi^{(s)}_{\nu\mu_2\dots\mu_s}\partial_\lambda\varphi^{(s)\lambda\mu_2\dots\mu_s}\non \\
    &+\frac{s(s-1)(d+2s-6)}{2(d+2s-4)}\partial_\mu\partial_\nu\varphi^{(s)\mu\nu\mu_3\dots\mu_{s}}\varphi^{(s-2)}_{\mu_3\dots\mu_{s}}\non \\
    &-\frac{s(s-1)(d+2s-6)(d+2s-5)}{4(d+2s-4)^2}\varphi^{(s-2)\mu_3\dots\mu_{s}}\left(\square-\frac{d+2s-4}{d+2s-6}m^2\right)\varphi^{(s-2)}_{\mu_3\dots\mu_{s}} \non \\
    & +\frac{s(s-1)(s-2)(d+2s-6)(d+2s-8)}{8(d+2s-4)^2}\partial_\mu\varphi^{(s-2)\mu\mu_4\dots\mu_{s}}\partial^\nu\varphi^{(s-2)}_{\nu\mu_4\dots\mu_{s}}\non \\
    &-\sum^s_{q=3}\left(\frac{(s-q+2)(s-q+1)(d+2s-2q-2)}{4(d+2s-2q)}\right)
    \label{d-dimSH}
    \\
    & \times\left[\frac{(d+2s-2q-1)}{2(d+2s-2q)}\varphi^{(s-q)\mu_{q+1}\dots\mu_s}\left(\square-\frac{(d+2s-2q)q(d+2s-q-3)}{2(d+2s-2q-2)(d+2s-2q-1)}m^2\right)\varphi^{(s-q)}_{\mu_{q+1}\dots\mu_s}\right.\non \\
    & \hspace{20pt}-\frac{(s-q)(d+2s-2q-4)}{2(d+2s-2q)}\partial_\mu\varphi^{(s-q)\mu\mu_{q+2}\dots\mu_s}\partial^\nu\varphi^{(s-q)}_{\nu\mu_{q+2}\dots\mu_s}\non \\
    & \left.\hspace{20pt}-\sqrt{\frac{(s-q+3)(q-2)(d+2s-q-1)}{(d+2s-2q+2)}}m\varphi^{(s-q)\mu_{q+1}\dots\mu_s}\partial^\mu\varphi^{(s-q+1)}_{\mu\mu_{q+1}\dots\mu_s}\right]~.\non 
\end{align}
This Lagrangian defines the Singh-Hagen model in $d$ dimensions, which has so far been described in the literature only in the $d=4$ case \cite {Singh:1974}, see below.

Let us analyse the equation of motion corresponding to \eqref{d-dimSH}.
It is useful to adopt the  Singh-Hagen notation 
$\{ \dots\}_{\rm S.T.}$, which denotes the symmetric and traceless component of the term within the brackets.
The equation of motion for the field $\varphi^{(s)}$ is
\bsubeq
\begin{align}
    &(-\square+m^2)\varphi^{(s)}_{\mu_1\dots\mu_s}+s\partial^\nu\{\partial_{\mu_1}\varphi^{(s)}_{\nu\mu_2\dots \mu_s}\}_{\rm S.T.}=\frac{s(s-1)(d+2s-6)}{2(d+2s-4)}\{\partial_{\mu_1}\partial_{\mu_2}\varphi^{(s-2)}_{\mu_3\dots\mu_s}\}_{\rm S.T.}\label{SH d-dim EoM (s)}~.
 \end{align} 
The equation of motion for the field $\varphi^{(s-2)}$ is    
\begin{align}
  &\frac{s(s-1)(d+2s-6)(d+2s-5)}{(d+2s-4)^2}\left(\square-\frac{d+2s-4}{d+2s-6}m^2\right)\varphi^{(s-2)}_{\mu_3\dots\mu_{s}}\non \\
    &+\frac{s(s-1)(s-2)(d+2s-6)(d+2s-8)}{2(d+2s-4)^2}\partial^\nu\{\partial_{\mu_3}\varphi^{(s-2)}_{\nu\mu_4\dots\mu_s}\}_{\rm S.T.} \label{SH d-dim EoM (s-2)}\\
    &+\frac{(s-1)(s-2)(d+2s-4)\sqrt{s}}{2(d+2s-6)}m\{\partial_{\mu_3}\varphi^{(s-3)}_{\mu_4\dots\mu_s}\}_{\rm S.T.}
    =\frac{s(s-1)(d+2s-6)}{(d+2s-4)}\partial^\nu\partial^\lambda\varphi^{(s)}_{\nu\lambda\mu_3\dots\mu_s}~. \non
\end{align}
Finally, the equation of motion for the fields $\varphi^{(s-q)}$, with $3\leq q \leq s$, are the following:
\begin{align}
    & \qquad \left(\frac{(s-q+2)(s-q+1)(d+2s-2q-2)}{4(d+2s-2q)}\right)\non \\
    & \times\left[\frac{(d+2s-2q-1)}{(d+2s-2q)}\left(\square-\frac{(d+2s-2q)q(d+2s-q-3)}{2(d+2s-2q-2)(d+2s-2q-1)}m^2\right)\varphi^{(s-q)}_{\mu_{q+1}\dots \mu_s}\right.\non \\
    & \hspace{20pt}+\frac{(s-q)(d+2s-2q-4)}{(d+2s-2q)}\partial^\nu\{\partial_{\mu_{q+1}}\varphi^{(s-q)}_{\nu\mu_{q+2}\dots\mu_s}\}_{\rm S.T.}\label{SH d-dim EoM (s-q)} \\
    & \left.\hspace{20pt}-\sqrt{\frac{(s-q+3)(q-2)(d+2s-q-1)}{(d+2s-2q+2)}}m\partial^\mu\varphi^{(s-q+1)}_{\mu\mu_{q+1}\dots\mu_s}\right] \non\\
    =&  -\left(\frac{(s-q+3)(s-q+2)(d+2s-2q)}{4(d+2s-2q+2)}     \right) \non \\
&\times     \sqrt{\frac{(s-q+4)(q-3)(d+2s-q)}{(d+2s-2q)}}m\{\partial_{\mu_{q+1}}\varphi^{(s-q-1)}_{\mu_{q+2}\dots\mu_s}\}_{\rm S.T.}~.\non
\end{align}
\esubeq
After some algebra, it may be seen that these equations yield $\varphi^{(s-q)}=0$ for $s\geq q\geq 2$, while 
the field $\varphi^{(s)}$ obeys the Fierz-Pauli equations
 \eqref{Fierz Pauli}. As a result, the number of on-shell degrees of freedom is 
 \bea
   n(d,s)  =   \frac{2s+d-3}{d-3}\binom{d+s-4}{s}~,
   \label{3.11}
\eea
which reduces to $2s+1$ degrees of freedom in four spacetime dimensions.  In three spacetime dimensions, 
\eqref{3.11} should be replaced with $n(3,s) =2$. 

As an instructive check, it is worth 
comparing \eqref{d-dimSH} directly with the Singh-Hagen model
 \cite {Singh:1974} formulated in ${\mathbb M}^4$.
Choosing $d=4$ in \eqref{d-dimSH} yields
\begin{align}
    \mathcal{L} = &\frac{1}{2}\varphi^{(s)\mu_1\dots\mu_s}(\square-m^2)\varphi^{(s)}_{\mu_{1}\dots\mu_{s}}+\frac{1}{2}s\partial^\nu\varphi^{(s)}_{\nu\mu_2\dots\mu_s}\partial_\lambda\varphi^{(s)\lambda\mu_2\dots\mu_s}+\frac{(s-1)^2}{2}\partial_\mu\partial_\nu\varphi^{(s)\mu\nu\mu_3\dots\mu_{s}}\varphi^{(s-2)}_{\mu_3\dots\mu_{s}}\non \\
    &-\frac{(s-1)^2(2s-1)}{8s}\varphi^{(s-2)\mu_3\dots\mu_{s}}\left(\square-\frac{s}{s-1}m^2\right)\varphi^{(s-2)}_{\mu_3\dots\mu_{s}}\non \\
    & +\frac{(s-1)^2(s-2)^2}{8s}\partial_\mu\varphi^{(s-2)\mu\mu_4\dots\mu_{s}}\partial^\nu\varphi^{(s-2)}_{\nu\mu_4\dots\mu_{s}}\\
    &-\sum^s_{q=3}\left[\frac{(s-q+1)^2(2s-2q+3)}{8(s-q+2)}\varphi^{(s-q)\mu_{q+1}\dots\mu_s}\left(\square-\frac{(s-q+2)k(2s-q+1)}{2(s-q+1)(2s-2q+3)}m^2\right)\varphi^{(s-q)}_{\mu_{q+1}\dots\mu_s}\right.\non \\
    & \hspace{30pt}-\frac{(s-q+1)^2(s-q)^2}{8(s-q+2)}\partial_\mu\varphi^{(s-q)\mu\mu_{q+2}\dots\mu_s}\partial^\nu\varphi^{(s-q)}_{\nu\mu_{q+2}\dots\mu_s}\non \\
    & \left.\hspace{30pt}-\frac{(s-q+1)^2}{4}\sqrt{\frac{(q-2)(2s-q+3)}{2}}m\varphi^{(s-q)\mu_{q+1}\dots\mu_s}\partial^\mu\varphi^{(s-q+1)}_{\mu\mu_{q+1}\dots\mu_s}\right]~.\non 
\end{align}
Let us rescale the fields $\varphi^{(s-2)}$ and $\varphi^{(s-q)}$, with $3\leq q\leq s$, as follows:
\bsubeq
\bea
    \varphi^{(s-2)}_{\mu_{3}\dots\mu_s}&\longrightarrow &\frac{2s}{2s-1}\varphi^{(s-2)}_{\mu_{3}\dots\mu_s}~, \\
    \varphi^{(s-q)}_{\mu_{q+1}\dots\mu_s}&\longrightarrow& \frac{2(s-1)}{(s-q+1)}\sqrt{\frac{s(s-q+2)}{(2s-1)(2s-2q+3)}}
    \non\\
    &&\times\left(\prod^{q-1}_{j=2}\sqrt{\frac{(j-1)(s-j)^2(s-j+2)(2s-j+2)}{2(s-j+1)(2s-2j+1)(2s-2j+3)}}\right)\varphi^{(s-q)}_{\mu_{q+1}\dots\mu_s}~.
\eea
\esubeq
This yields
\begin{align}
    \mathcal{L} = &\frac{1}{2}\varphi^{(s)\mu_1\dots\mu_s}(\square-m^2)\varphi^{(s)}_{\mu_{1}\dots\mu_{s}}+\frac{1}{2}s\partial^\nu\varphi^{(s)}_{\nu\mu_2\dots\mu_s}\partial_\lambda\varphi^{(s)\lambda\mu_2\dots\mu_s}\non \\
    &+\frac{s(s-1)^2}{2s-1}\left\{\partial_\mu\partial_\nu\varphi^{(s)\mu\nu\mu_3\dots\mu_{s}}\varphi^{(s-2)}_{\mu_3\dots\mu_{s}}-\frac{1}{2}\varphi^{(s-2)\mu_3\dots\mu_{s}}\left(\square-\frac{s}{s-1}m^2\right)\varphi^{(s-2)}_{\mu_3\dots\mu_{s}}\right.\non \\
    & +\frac{(s-2)^2}{2(2s-1)}\partial_\mu\varphi^{(s-2)\mu\mu_4\dots\mu_{s}}\partial^\nu\varphi^{(s-2)}_{\nu\mu_4\dots\mu_{s}}\\
    &-\sum^s_{q=3}\left(\prod^{q-1}_{j=2}\frac{(j-1)(s-j)^2(s-j+2)(2s-j+2)}{2(s-j+1)(2s-2j+1)(2s-2j+3)}\right)\non \\
    & \times\left[\frac{1}{2}\varphi^{(s-q)\mu_{q+1}\dots\mu_s}\left(\square-\frac{q(s-q+2)(2s-q+1)}{2(s-q+1)(2s-2q+3)}m^2\right)\varphi^{(s-q)}_{\mu_{q+1}\dots\mu_s}\right.\non \\
    & \hspace{15pt} \left.\left.-\frac{(s-q)^2}{2(2s-2q+3)}\partial_\mu\varphi^{(s-q)\mu\mu_{q+2}\dots\mu_s}\partial^\nu\varphi^{(s-q)}_{\nu\mu_{q+2}\dots\mu_s}-m\varphi^{(s-q)\mu_{q+1}\dots\mu_s}\partial^\mu\varphi^{(s-q+1)}_{\mu\mu_{q+1}\dots\mu_s}\right]\right\}~,\non 
\end{align}
which is exactly the Lagrangian derived by  Singh and Hagen  \cite {Singh:1974}.


\section{Quantisation: Spin $s=2$}
\label{Spin 2}

The Zinoviev theory \eqref{Full Action for spin s} 
is an irreducible gauge theory 
 (following the terminology of the Batalin-Vilkovisky formalism \cite{BV})
and  can be quantised \`a la Faddeev and Popov \cite{Faddeev:1967fc}.
Before considering its quantisation 
in the general $s\geq 2$ case, 
it is worth investigating how the process is carried out for the simplest $s=2$ and $s=3$ values. 

In the spin $s=2$ case, the action (\ref{Full Action for spin s}) is
\begin{align}
        S  = & \int \rd^dx \,\left\{-\frac{1}{2}\partial^{\mu}\phi^{(2)\nu\lambda}\partial_\mu\phi^{(2)}_{\nu\lambda}+\partial_\mu\phi^{(2)\mu\lambda}\partial^\nu\phi^{(2)}_{\nu\lambda}+\frac{1}{2}\partial^\mu\Tilde{\phi}^{(2)}\partial_\mu\Tilde{\phi}^{(2)}\right.\non \\
        & \hspace{40pt}+\partial^\mu\partial^\nu\phi^{(2)}_{\mu\nu}\Tilde{\phi}^{(2)}-\frac{1}{2}\partial^\mu\phi^{(1)\nu}\partial_\mu\phi^{(1)}_\nu+\frac{1}{2}\partial^\mu\phi^{(1)}_\mu\partial^\nu\phi^{(1)}_\nu\non \\
        & \hspace{40pt}-\frac{1}{2}\partial^\mu\phi^{(0)}\partial_\mu\phi^{(0)}+a_2\phi^{(1)\nu}\partial^\mu\phi^{(2)}_{\mu\nu}+b_2\Tilde{\phi}^{(2)}\partial^\mu\phi^{(1)}_\mu\label{Spin 2 Action}\\
        & \hspace{40pt}+d_2\phi^{(2)\mu\nu}\phi^{(2)}_{\mu\nu}+e_2(\Tilde{\phi}^{(2)})^2+a_1\phi^{(0)}\partial^\mu\phi^{(1)}_\mu\non \\
        & \hspace{40pt}\left.-f_2\Tilde{\phi}^{(2)}\phi^{(0)}+d_1\phi^{(1)\mu}\phi^{(1)}_\mu+d_0(\phi^{(0)})^2\right\}~,\non 
\end{align}
and the gauge transformation (\ref{k^th Gauge Transformation}) reads
\bsubeq
    \label{Spin 2 Gauge Transformation}
    \begin{align}
        \delta\phi^{(2)}_{\mu\nu} & = \frac{1}{2}\left(\partial_\mu\xi^{(1)}_\nu+\partial_\nu\xi^{(1)}_\mu\right)-\beta_2\eta_{\mu\nu}\xi^{(0)}~,\label{spin 2 gauge transformation 1}\\
        \noalign{\vspace{10pt}} \delta\phi^{(1)}_\mu & = \partial_\mu\xi^{(0)}+\alpha_1\xi^{(1)}_\mu~,\label{spin 2 gauge transformation 2}\\
        \noalign{\vspace{10pt}} \delta\phi^{(0)} & = \alpha_0\xi^{(0)}~.\label{spin 2 gauge transformation 3}
    \end{align} 
\esubeq

To carry out the Faddeev-Popov scheme, suitable gauge conditions are required. We choose the following gauge-fixing functions:
\bsubeq
    \begin{align}
        &\Xi^{(1)}_{\mu}
        -\chi^{(1)}_{\mu} 
        =2\partial^\nu\phi^{(2)}_{\mu\nu}-\partial_\mu\Tilde{\phi}^{(2)}-m\sqrt{2}\phi^{(1)}_\mu-\chi^{(1)}_\mu ~,\label{Spin 2 Gauge Fixing 3}\\
        \noalign{\vspace{10pt}} &\Xi^{(0)}
        -\chi^{(0)} 
        = \partial^\mu\phi^{(1)}_\mu
        -\frac{m}{\sqrt{2}}\Tilde{\phi}^{(2)}
        -\frac{2d-2}{d-2}\frac{m^2}{\alpha_0}\phi^{(0)}-\chi^{(0)} \label{Spin 2 Gauge Fixing 4}~,
    \end{align}
\esubeq
where $\chi^{(1)}$ and $\chi^{(0)}$ are background fields. The explicit expressions for $\Xi^{(1)}$ and $\Xi^{(0)}$  have been chosen so that their gauge variations are
\bsubeq
    \begin{align}
        \delta\Xi^{(1)}_\mu=(\square-m^2)\xi^{(1)}_\mu \label{Xi 1 Varied}~,\\
        \noalign{\vspace{10pt}} \delta\Xi^{(0)}=(\square-m^2)\xi^{(0)}\label{Xi 0 Varied}~.
    \end{align}
\esubeq
With these, we have the partition function 
\begin{equation}
    Z^{(2)} = \int\mathcal{D}(\phi;2)\Delta^{(1)}\Delta^{(0)}\delta\left[\Xi^{(1)}_\mu-\chi^{(1)}_\mu\right]\delta\left[\Xi^{(0)}-\chi^{(0)}\right]\re^{\ri S} \label{Spin 2 Partition Function 1}~,
\end{equation}
where the objects $\Delta^{(1)}$ and $\Delta^{(0)}$ are 
the Faddeev-Popov determinants, 
\bsubeq
    \begin{align}
        \Delta^{(0)} &= \det \left(\frac{\delta \Xi^{(0)}(x)}{\delta\xi^{(0)}(x')}\right) = \det \left[(\square-m^2)\delta^d(x-x')\right]~,\label{Spin 2 First Delta 1}\\
        \noalign{\vspace{10pt}} \Delta^{(1)} &= \det\left( \frac{\delta \Xi^{(1)}_\mu(x)}{\d \xi^{(1)}_\nu(x')}\right) = \det[ \d_\m{}^\n(\square-m^2)\delta^d(x-x')]~. \label{Spin 2 Second Delta 1}
    \end{align}
\esubeq

Since the partition function \eqref{Spin 2 Partition Function 1} is independent of the background fields $\chi^{(1)}$ and $\chi^{(0)}$, we can average over them with a convenient weight of the form
\bea
    \exp\left\{-\frac{\ri}{2}\int \rd^dx \,\left[\frac{\chi^{(1)\mu}\chi^{(1)}_\mu}{\omega_1}+\frac{(\chi^{(0)})^2}{\omega_0}\right]\right\}~,
\eea
where $\omega_1$ and $\omega_0$ are constants that will be chosen in such a way as to diagonalise the action as well as cause all divergence terms to vanish. 
Doing so, (\ref{Spin 2 Partition Function 1}) becomes
\begin{align}
        Z^{(2)} = & \int\mathcal{D}(\phi;2)\Delta^{(1)}\Delta^{(0)}\non \\
        & \times\exp\left\{-\frac{\ri}{2}\int \rd^dx \,\left[\frac{1}{\omega_1}\left(2\partial^\nu\phi^{(2)}_{\mu\nu}-\partial_\mu\Tilde{\phi}^{(2)}-m\sqrt{2}\phi^{(1)}_\mu\right)^2\right.\right.
        \label{Spin 2 Partition Function 2}\\
        & \hspace{100pt}\left.\left.+\frac{1}{\omega_0}\left(\partial^\mu\phi^{(1)}_\mu
        -\frac{m}{\sqrt{2}}\Tilde{\phi}^{(2)}
        -\frac{2d-2}{d-2}\frac{m^2}{\alpha_0}\phi^{(0)}\right)^2\right]\right\}\re^{\ri S} ~.\non
\end{align}
Substituting the classical action (\ref{Spin 2 Action}) in \eqref{Spin 2 Partition Function 2} yields
\begin{align}
        Z^{(2)} = & \int\mathcal{D}(\phi;2)\Delta^{(1)}\Delta^{(0)}\exp\left\{\ri\int \rd^dx \,\left[-\frac{1}{2}\partial^{\mu}\phi^{(2)\nu\lambda}\partial_\mu\phi^{(2)}_{\nu\lambda}+\left(1-\frac{2}{\omega_1}\right)\partial_\mu\phi^{(2)\mu\lambda}\partial^\nu\phi^{(2)}_{\nu\lambda}\right.\right.\non \\
        & +\frac{1}{2}\left(1-\frac{1}{\omega_1}\right)\partial^\mu\Tilde{\phi}^{(2)}\partial_\mu\Tilde{\phi}^{(2)}+\left(1-\frac{2}{\omega_1}\right)\partial^\mu\partial^\nu\phi^{(2)}_{\mu\nu}\Tilde{\phi}^{(2)}-\frac{1}{2}\partial^\mu\phi^{(1)\nu}\partial_\mu\phi^{(1)}_\nu\non \\
        & +\left(d_1-\frac{m^2}{\omega_1}\right)\phi^{(1)\mu}\phi^{(1)}_\mu+\frac{1}{2}\left(1-\frac{1}{\omega_0}\right)\partial^\mu\phi^{(1)}_\mu\partial^\nu\phi^{(1)}_\nu-\frac{1}{2}\partial^\mu\phi^{(0)}\partial_\mu\phi^{(0)}\non \\
        & +\left(a_2+\frac{2\sqrt{2}m}{\omega_1}\right)\phi^{(1)\nu}\partial^\mu\phi^{(2)}_{\mu\nu}+\left(b_2+\frac{m}{\sqrt{2}\omega_0}+\frac{\sqrt{2}m}{\omega_1}\right)\Tilde{\phi}^{(2)}\partial^\mu\phi^{(1)}_\mu\label{Spin 2 Partition Function 3}\\
        & +d_2\phi^{(2)\mu\nu}\phi^{(2)}_{\mu\nu}+\left(e_2-\frac{m^2}{4\omega_0}\right)(\Tilde{\phi}^{(2)})^2-\left(f_2+\frac{\sqrt{2}(d-1)}{d-2}\frac{m^3}{\alpha_0\omega_0}\right)\Tilde{\phi}^{(2)}\phi^{(0)}\non \\
        & \left.\left.+\left(a_1+\frac{2d-2}{d-2}\frac{m^2}{\alpha_0\omega_0}\right)\phi^{(0)}\partial^\mu\phi^{(1)}_\mu+\left(d_0-\frac{(2d-2)^2}{2(d-2)^2}\frac{m^4}{(\alpha_0)^2\omega_0}\right)(\phi^{(0)})^2\right]\right\} \non ~.
\end{align}
To get our desired result, we choose 
\bea
    \omega_1 = 2~, \qquad
    \omega_0 = 1~.
\eea
With such a choice, 
making use of (\ref{Zinoviev Coefficients}) and (\ref{Alpha k}) gives
\begin{align}
        Z^{(2)} = & \int\mathcal{D}(\phi;2)\Delta^{(1)}\Delta^{(0)}\exp\left\{\ri\int \rd^dx \,\left[\frac{1}{2}\phi^{(2)\mu\nu}\left(\square-m^2\right)\phi^{(2)}_{\mu\nu}-\frac{1}{4}\Tilde{\phi}^{(2)}\left(\square-m^2\right)\Tilde{\phi}^{(2)}\right.\right.\non \\
        & \left.\left.+\frac{1}{2}\phi^{(1)\mu}\left(\square-m^2\right)\phi^{(1)}_\mu+\frac{1}{2}\phi^{(0)}\left(\square-m^2\right)\phi^{(0)}\right]\right\} \label{Spin 2 Partition Function 6}~.
 \end{align}

It only remains to recast 
the ghost contributions in terms of path integrals.
In accordance with (\ref{Determinant of M}) we rewrite (\ref{Spin 2 First Delta 1}) and (\ref{Spin 2 Second Delta 1}) as follows:
\begin{subequations}
\bea
    \Delta^{(0)} &=& \int \mathcal{D}\overline{\psi}^{(0)}\mathcal{D}\psi^{(0)}\exp\left[-\ri\int \rd^dx \,\:\overline{\psi}^{(0)}(\square-m^2)\psi^{(0)}\right] \label{Spin 2 First Delta 2}~,\\
    \Delta^{(1)} &=& \int \mathcal{D}\overline{\psi}^{(1)}\mathcal{D}\psi^{(1)}\exp\left[-\ri\int \rd^dx \,\:\overline{\psi}^{(1)\mu}(\square-m^2)\psi^{(1)}_\mu\right] \label{Spin 2 Second Delta 2}~.
\eea
\end{subequations}

With these, (\ref{Spin 2 Partition Function 6}) becomes
\begin{align}
        Z^{(2)} = & \int\mathcal{D}(\phi,\psi;2)\exp\left\{\ri\int \rd^dx \,\left[\frac{1}{2}\phi^{(2)\mu\nu}\left(\square-m^2\right)\phi^{(2)}_{\mu\nu}-\frac{1}{4}\Tilde{\phi}^{(2)}\left(\square-m^2\right)\Tilde{\phi}^{(2)}\right.\right.\non\\
        & -\overline{\psi}^{(1)\mu}(\square-m^2)\psi^{(1)}_\mu+\frac{1}{2}\phi^{(1)\mu}\left(\square-m^2\right)\phi^{(1)}_\mu-\overline{\psi}^{(0)}(\square-m^2)\psi^{(0)}\label{Spin 2 Partition Function 7}\\
        & \left.\left.+\frac{1}{2}\phi^{(0)}\left(\square-m^2\right)\phi^{(0)}\right]\right\} ~,\non
\end{align}
which is the fully diagonalised partition function for the spin-2 case.


\section{Quantisation: Spin $s=3$}
\label{Spin 3}

A similar procedure is carried out for the spin-3 field. The corresponding classical action is
\begin{align}
        S  = & \int \rd^dx \,\left\{-\frac{1}{2}\partial^\mu\phi^{(3)\nu\lambda\rho}\partial_\mu\phi^{(3)}_{\nu\lambda\rho}+\frac{3}{2}\partial_\mu\phi^{(3)\mu\lambda\rho}\partial^\nu\phi^{(3)}_{\nu\lambda\rho}+\frac{3}{2}\partial^\mu\Tilde{\phi}^{(3)\nu}\partial_\mu\Tilde{\phi}^{(3)}_\nu\right.\non \\
        & \hspace{40pt} +3\partial_\mu\partial_\nu\phi^{(3)\mu\nu\lambda}\Tilde{\phi}^{(3)}_{\lambda}+\frac{3}{4}\partial_\mu\Tilde{\phi}^{(3)\mu}\partial_\nu\Tilde{\phi}^{(3)\nu}+a_3\phi^{(2)\nu\lambda}\partial^\mu\phi^{(3)}_{\mu\nu\lambda}\non \\
        & \hspace{40pt} +b_3m\Tilde{\phi}^{(3)\nu}\partial^\mu\phi^{(2)}_{\mu\nu}+c_3\partial_\mu\Tilde{\phi}^{(3)\mu}\Tilde{\phi}^{(2)}+d_3\phi^{(3)\mu\nu\lambda}\phi^{(3)}_{\mu\nu\lambda}+e_3\Tilde{\phi}^{(3)\mu}\Tilde{\phi}^{(3)}_{\mu}\non \\
        & \hspace{40pt} -f_3\Tilde{\phi}^{(3)\mu}\phi^{(1)}_{\mu}-\frac{1}{2}\partial^{\mu}\phi^{(2)\nu\lambda}\partial_\mu\phi^{(2)}_{\nu\lambda}+\partial_\mu\phi^{(2)\mu\lambda}\partial^\nu\phi^{(2)}_{\nu\lambda}
        \label{Spin 3 Action}\\
        & \hspace{40pt} +\frac{1}{2}\partial^\mu\Tilde{\phi}^{(2)}\partial_\mu\Tilde{\phi}^{(2)}+\partial^\mu\partial^\nu\phi^{(2)}_{\mu\nu}\Tilde{\phi}^{(2)}-\frac{1}{2}\partial^\mu\phi^{(1)\nu}\partial_\mu\phi^{(1)}_\nu\non \\
        & \hspace{40pt} +\frac{1}{2}\partial^\mu\phi^{(1)}_\mu\partial_\nu\phi^{(1)}_\nu-\frac{1}{2}\partial^\mu\phi^{(0)}\partial_\mu\phi^{(0)}+a_2\phi^{(1)\nu}\partial^\mu\phi^{(2)}_{\mu\nu}+b_2\Tilde{\phi}^{(2)}\partial^\mu\phi^{(1)}_\mu\non \\
        & \hspace{40pt} +d_2\phi^{(2)\mu\nu}\phi^{(2)}_{\mu\nu}+e_2(\Tilde{\phi}^{(2)})^2-f_2\Tilde{\phi}^{(2)}\phi^{(0)}+a_1m\phi^{(0)}\partial^\mu\phi^{(1)}_\mu+d_1\phi^{(1)\mu}\phi^{(1)}_\mu\non \\
        & \hspace{40pt} \left.+d_0(\phi^{(0)})^2\right\} ~.
        \non 
 \end{align}
It is invariant under the gauge transformations
\bsubeq
\label{Spin 3 Gauge Transformations}
    \begin{align}
        \delta\phi^{(3)}_{\mu\nu\lambda} &= \frac{1}{3}\left(\partial_\mu\xi^{(2)}_{\nu\lambda}+\partial_\nu\xi^{(2)}_{\mu\lambda}+\partial_\lambda\xi^{(2)}_{\mu\nu}\right)-\beta_3\left(\eta_{\mu\nu}\xi^{(1)}_{\lambda}+\eta_{\lambda\mu}\xi^{(1)}_{\nu}+\eta_{\nu\lambda}\xi^{(1)}_{\mu}\right)~, \label{Spin 3 Gauge Transformation 1}\\
        \noalign{\vspace{10pt}}\delta\phi^{(2)}_{\mu\nu} &= \frac{1}{2}\left(\partial_\mu\xi^{(1)}_\nu+\partial_\nu\xi^{(1)}_\mu\right)+\alpha_2\xi^{(2)}_{\mu\nu}-\beta_2\eta_{\mu\nu}\xi^{(0)}~, \label{Spin 3 Gauge Transformation 2}\\
        \noalign{\vspace{10pt}} \delta\phi^{(1)}_\mu &= \partial_\mu\xi^{(0)}+\alpha_1\xi^{(1)}_\mu~, \label{Spin 3 Gauge Transformation 3}\\
        \noalign{\vspace{10pt}} \delta\phi^{(0)} &= \alpha_0\xi^{(0)}~. \label{Spin 3 Gauge Transformation 4}
    \end{align} 
\esubeq

To quantise the theory,  we choose  the following gauge-fixing  functions:
\bsubeq \label{5.3}
    \begin{align}
        &\Xi^{(2)}_{\mu\nu}-\chi^{(2)}_{\mu\nu} = 3\partial^\lambda\phi^{(3)}_{\mu\nu\lambda} - 3\partial_{(\mu}\Tilde{\phi}^{(3)}_{\nu)} - \sqrt{3}m\phi^{(2)}_{\mu\nu} + \frac{\sqrt{3}}{d}\eta_{\mu\nu}m\Tilde{\phi}^{(2)}-\chi^{(2)}_{\mu\nu} ~,\label{Spin 3 Gauge Fixing 7}\\
        \noalign{\vspace{10pt}} &\Xi^{(1)}_{\mu}-\chi^{(1)}_{\mu} =  2\partial^\nu\phi^{(2)}_{\mu\nu}
        -\sqrt{3}m\Tilde{\phi}^{(3)}_\mu  - \partial_\mu\Tilde{\phi}^{(2)} - 2m\sqrt{\frac{d+1}{d}}\phi^{(1)}_\mu-\chi^{(1)}_\mu ~,\label{Spin 3 Gauge Fixing 8}\\
        \noalign{\vspace{10pt}} &\Xi^{(0)}-\chi^{(0)} = \partial^\mu\phi^{(1)}_\mu  -m\sqrt{\frac{d+1}{d}}\Tilde{\phi}^{(2)} - m\sqrt{\frac{3d}{d-2}}\phi^{(0)}-\chi^{(0)}~,\label{Spin 3 Gauge Fixing 9}
    \end{align}
\esubeq
where $\chi^{(2)}_{\mu\nu} $, $\chi^{(1)}_{\mu}$ and $\chi^{(0)} $ are background fields. Here both  $\Xi^{(2)}_{\mu\nu}$  and $\c^{(2)}_{\mu\nu}$ are symmetric and traceless. 
The gauge-fixing functions \eqref{5.3} have been chosen so that $\Xi^{(2)} _{\mu\nu} $ 
varies as
\bea
    \delta\Xi^{(2)} _{\mu\nu} = (\square-m^2)\xi^{(2)}_{\mu\nu}~,
\eea
while the gauge variations $ \d \Xi^{(1)}$ and $\d \Xi^{(0)}$ are given by 
(\ref{Xi 1 Varied}) and (\ref{Xi 0 Varied}), respectively.
The partition function is then
\begin{equation}
    Z^{(3)} = \int\mathcal{D}(\phi;3)\Delta^{(2)}\Delta^{(1)}\Delta^{(0)}\delta\left[\Xi^{(2)}_{\mu\nu}-\chi^{(2)}_{\mu\nu}\right]\delta\left[\Xi^{(1)}_\mu-\chi^{(1)}_\mu\right]\delta\left[\Xi^{(0)}-\chi^{(0)}\right]\re^{\ri S}~, \label{Spin 3 Partition Function 1}
\end{equation}
where once again $\Delta^{(0)}$ and $\Delta^{(1)}$ are ghost contributions of spin 0 and 1 while $\Delta^{(2)}$ is the ghost contribution of spin 2. 

Since the partition function \eqref{Spin 3 Partition Function 1} is independent of the background fields $\chi^{(2)}$, $\chi^{(1)}$ and $\chi^{(0)}$, we can average over them with a convenient weight of the form
\bea
    \exp\left\{-\frac{\ri}{2}\int \rd^dx \,\Big[\frac{\chi^{(2)\mu\nu}\chi^{(2)}_{\mu\nu}}{\omega_2}+\frac{\chi^{(1)\mu}\chi^{(1)}_\mu}{\omega_1}+\frac{(\chi^{(0)})^2}{\omega_0}\Big]\right\}~.
    \label{5.6}
\eea
Upon doing so, the relation (\ref{Spin 3 Partition Function 1}) turns into
\bea
        Z^{(3)} &= & \int\mathcal{D}(\phi;3)\Delta^{(2)}\Delta^{(1)}\Delta^{(0)}\displaystyle \exp
        \left\{-\frac{\ri}{2}\int \rd^dx \,\right.\non \\
        && \times
        \bigg[
        \frac{1} {\omega_2} \Big(3\partial_\lambda\phi^{(3)\mu\nu\lambda} - 3\partial^{(\mu}\Tilde{\phi}^{(3)\nu)} - \sqrt{3}m\phi^{(2)\mu\nu} + \frac{\sqrt{3}}{d}\eta^{\mu\nu}m\Tilde{\phi}^{(2)}\Big)^2 \non \\
        && +\frac{1}{\omega_1}  \Big(2\partial^\nu\phi^{(2)}_{\mu\nu} 
        -\sqrt{3}m\Tilde{\phi}^{(3)}_\mu 
        - \partial_\mu\Tilde{\phi}^{(2)} 
        - 2m\sqrt{\frac{d+1}{d}}\phi^{(1)}_\mu\Big)^2 
        \label{Spin 3 Partition Function 2}\\
        && \left.
        +\frac{1} {\omega_0} \Big( \partial^\mu\phi^{(1)}_\mu
        -m\sqrt{\frac{d+1}{d}}\Tilde{\phi}^{(2)}  - m\sqrt{\frac{3d}{d-2}}\phi^{(0)}\Big)^2  \bigg]\right\}\re^{\ri S}\non ~.
\eea
This is then combined with (\ref{Spin 3 Action}) to get
\bea
        Z^{(3)} &= & \int\mathcal{D}(\phi;3)\Delta^{(2)}\Delta^{(1)}\Delta^{(0)}\displaystyle \exp\left\{\ri\int \rd^dx \,\left[-\frac{1}{2}\partial^\mu\phi^{(3)\nu\lambda\rho}\partial_\mu\phi^{(3)}_{\nu\lambda\rho}
        +\left(\frac{3}{2}-\frac{9}{2\omega_2}\right)
        \partial_\mu\phi^{(3)\mu\lambda\rho}\partial^\nu\phi^{(3)}_{\nu\lambda\rho}\right.\right.\non \\
        && +\left(\frac{3}{2}-\frac{9}{4\omega_2}\right)\partial^\mu\Tilde{\phi}^{(3)\nu}\partial_\mu\Tilde{\phi}^{(3)}_\nu+\left(3-\frac{9}{\omega_2}\right)\partial_\mu\partial_\nu\phi^{(3)\mu\nu\lambda}\Tilde{\phi}^{(3)}_{\lambda}+\left(\frac{3}{4}-\frac{9}{4\omega_2}\right)\partial_\mu\Tilde{\phi}^{(3)\mu}\partial_\nu\Tilde{\phi}^{(3)\nu}\non \\
        && +\left(a_3+\frac{3\sqrt{3}m}{\omega_2}\right)\phi^{(2)\nu\lambda}\partial^\mu\phi^{(3)}_{\mu\nu\lambda}+\left(b_3+\frac{2\sqrt{3}m}{\omega_1}+\frac{3\sqrt{3}m}{\omega_2}\right)\Tilde{\phi}^{(3)\nu}\partial^\mu\phi^{(2)}_{\mu\nu}\non \\
        && +\left(c_3+\frac{\sqrt{3}m}{\omega_1}\right)\partial_\mu\Tilde{\phi}^{(3)\mu}\Tilde{\phi}^{(2)}+\frac{m^2}{2}\phi^{(3)\mu\nu\lambda}\phi^{(3)}_{\mu\nu\lambda}+\left(e_3-\frac{3m^2}{2\omega_1}\right)\Tilde{\phi}^{(3)\mu}\Tilde{\phi}^{(3)}_{\mu}\non \\
        && -\left(f_3+\frac{2m^2}{\omega_1}\sqrt{\frac{3(d+1)}{d}}\right)\Tilde{\phi}^{(3)\mu}\phi^{(1)}_{\mu}-\frac{1}{2}\partial^{\mu}\phi^{(2)\nu\lambda}\partial_\mu\phi^{(2)}_{\nu\lambda}+\left(1-\frac{2}{\omega_1}\right)\partial_\mu\phi^{(2)\mu\lambda}\partial^\nu\phi^{(2)}_{\nu\lambda}\non \\
        && +\left(\frac{1}{2}-\frac{1}{2\omega_1}\right)\partial^\mu\Tilde{\phi}^{(2)}\partial_\mu\Tilde{\phi}^{(2)}+\left(1-\frac{2}{\omega_2}\right)\partial^\mu\partial^\nu\phi^{(2)}_{\mu\nu}\Tilde{\phi}^{(2)}-\frac{1}{2}\partial^\mu\phi^{(1)\nu}\partial_\mu\phi^{(1)}_\nu\label{Spin 3 Partition Function 3}\\
        && +\left(\frac{1}{2}-\frac{1}{2\omega_0}\right)\partial^\mu\phi^{(1)}_\mu\partial_\nu\phi^{(1)}_\nu+\frac{1}{2}\partial^\mu\phi^{(0)}\partial_\mu\phi^{(0)}+\left(a_2+\frac{4m}{\omega_1}\sqrt{\frac{d+1}{d}}\right)\phi^{(1)\nu}\partial^\mu\phi^{(2)}_{\mu\nu}\non \\
        && +\left(b_2+\frac{m}{\omega_0}\sqrt{\frac{d+1}{d}}+\frac{2m}{\omega_1}\sqrt{\frac{d+1}{d}}\right)\Tilde{\phi}^{(2)}\partial^\mu\phi^{(1)}_\mu-\frac{3}{2\omega_2}m^2\phi^{(2)\mu\nu}\phi^{(2)}_{\mu\nu}\non \\
        && +\left(e_2-\frac{3m^2}{2d\omega_2}-\frac{(d+1)m^2}{2d\omega_0}\right)(\Tilde{\phi}^{(2)})^2-\left(f_2+\frac{m^2}{\omega_0}\sqrt{\frac{3(d+1)}{d-2}}\right)m^2\Tilde{\phi}^{(2)}\phi^{(0)}\non \\
        && +\left(a_1+\frac{m}{\omega_0}\sqrt{\frac{3d}{d-2}}\right)\phi^{(0)}\partial^\mu\phi^{(1)}_\mu+\left(d_1-\frac{2(d+1)m^2}{d\omega_1}\right)\phi^{(1)\mu}\phi^{(1)}_\mu\non \\
        && \left.\left.+\left(d_0-\frac{3dm^2}{2(d-2)\omega_0}\right)(\phi^{(0)})^2\right]\right\}~.\non 
\eea
It may be seen that the coefficients in \eqref{5.6} that diagonalise the action and remove all terms with divergences (such as $\partial_\mu\phi^{(3)\mu\lambda\rho}\partial^\nu\phi^{(3)}_{\nu\lambda\rho}$) are:
\bea
    \omega_2 = 3~, \qquad \omega_1 = 2~, \qquad \omega_0 = 1~.
\eea
Taking into account the relations (\ref{Zinoviev Coefficients}) and integrating by parts, one arrives at
\bea
        Z^{(3)} &= & \int\mathcal{D}(\phi;3)\Delta^{(2)}\Delta^{(1)}\Delta^{(0)}\displaystyle \exp\left\{\ri\int \rd^dx \,\left[\frac{1}{2}\phi^{(3)\mu\nu\lambda}\left(\square-m^2\right)\phi^{(3)}_{\mu\nu\lambda}\right.\right.\non\\
        && -\frac{3}{4}\Tilde{\phi}^{(3)\mu}\left(\square-m^2\right)\Tilde{\phi}^{(3)}_{\mu}+\frac{1}{2}\phi^{(2)\mu\nu}\left(\square-m^2\right)\phi^{(2)}_{\mu\nu}-\frac{1}{4}\Tilde{\phi}^{(2)}\left(\square-m^2\right)\Tilde{\phi}^{(2)}\label{Spin 3 Partition Function 4}\\
        && \left.\left.+\frac{1}{2}\phi^{(1)\mu}\left(\square-m^2\right)\phi^{(1)}_{\mu}+\frac{1}{2}\phi^{(0)}\left(\square-m^2\right)\phi^{(0)}\right]\right\}~.\non
\eea

Finally, it only remains to massage 
the ghost contributions. 
The spin 0 and 1 ghost contributions are just (\ref{Spin 2 First Delta 2}) and (\ref{Spin 2 Second Delta 2}) as before. The spin 3 contribution is given by
\bea
    \Delta^{(2)} = \det\left(\frac{\delta \Xi^{(2)}_{\mu\nu} (x)}{\delta \xi^{(2)}_{\lambda\rho}(x')}\right) = \det \Big[ \hat{\d}_{\m \n}{}^{\l \r} (\square-m^2) \d^d(x-x')
      \Big] ~, \label{Spin 3 Third Delta 1}
\eea
where $ \hat{\d}_{\m \n}{}^{\l \r} $ is the Kronecker delta on the space of symmetric traceless second-rank tensors,
\bea
  \hat{\d}_{\m \n}{}^{\l \r} = \d_{(\m}{}^\l \d_{\n)}{}^\r - \frac{1}{d} \eta_{\m\n} \eta^{\l\r}~.
  \eea
Upon making use of (\ref{Determinant of M}), one obtains
\begin{equation}
    \Delta^{(2)} = \int \mathcal{D}\overline{\psi}^{(2)}\mathcal{D}\psi^{(2)}\exp\left[-\ri\int \rd^dx \,\:\overline{\psi}^{(2)\mu\nu}\left(\square-m^2\right)\psi^{(2)}_{\mu\nu}\right] \label{Third Delta 2}~.
\end{equation}
With this, the partition (\ref{Spin 3 Partition Function 4}) takes the form
\bea
        Z^{(3)} &= & \int\mathcal{D}(\phi,\psi;3)\displaystyle \exp\left\{\ri\int \rd^dx \,\left[\frac{1}{2}\phi^{(3)\mu\nu\lambda}\left(\square-m^2\right)\phi^{(3)}_{\mu\nu\lambda}-\frac{3}{4}\Tilde{\phi}^{(3)\mu}\left(\square-m^2\right)\Tilde{\phi}^{(3)}_{\mu}\right.\right.\non\\
        && -\overline{\psi}^{(2)\mu\nu}\left(\square-m^2\right)\psi^{(2)}_{\mu\nu}+\frac{1}{2}\phi^{(2)\mu\nu}\left(\square-m^2\right)\phi^{(2)}_{\mu\nu}-\frac{1}{4}\Tilde{\phi}^{(2)}\left(\square-m^2\right)\Tilde{\phi}^{(2)}\label{Spin 3 Partition Function 5}\\
        && -\overline{\psi}^{(1)\mu}(\square-m^2)\psi^{(1)}_\mu+\frac{1}{2}\phi^{(1)\mu}\left(\square-m^2\right)\phi^{(1)}_{\mu}-\overline{\psi}^{(0)}(\square-m^2)\psi^{(0)}\non\\
        && \left.\left.+\frac{1}{2}\phi^{(0)}\left(\square-m^2\right)\phi^{(0)}\right]\right\}~.\non
  \eea
It is seen that the gauge-fixed action is fully diagonalised.


\section{Quantisation: Arbitrary integer spin $s$}
\label{Spin s}

In the spin-$s$ case, the gauge-invariant action has the form (\ref{Full Action for spin s}). The corresponding  gauge freedom  is given by the transformations (\ref{k^th Gauge Transformation}) where $0\leq k \leq s$. 

To quantise the theory, we introduce the following symmetric and traceless gauge-fixing functions:  
\begin{align}
        \Xi^{(k)}_{\mu_1\dots\mu_k}-\chi^{(k)}_{\mu_1\dots\mu_k} &= (k+1)\partial^\mu\phi^{(k+1)}_{\mu\mu_1\dots\mu_{k}}
        -\frac{(k+2)(k+1)}{2}\alpha_{k+1}\Tilde{\phi}^{(k+2)}_{\mu_1\dots\mu_{k}}
        ~.\non\\
        &\hspace{15pt}-\frac{k(k+1)}{2}\partial_{(\mu_1}\Tilde{\phi}^{(k+1)}_{\mu_2\dots\mu_{k})}-\frac{(k+1)k(d+2k-4)}{2}\beta_{k+1}\phi^{(k)}_{\mu_1\dots\mu_{k}} \non\\
        &\hspace{15pt}+\frac{(k+1)k^2(k-1)}{4}\beta_{k+1}\eta_{(\mu_1\mu_2}\Tilde{\phi}^{(k)}_{\mu_3\dots\mu_{k})}-\chi^{(k)}_{\mu_1\dots\mu_k}
                \label{General kth Gauge Fixing 5}~,
 \end{align}
where $\chi^{(k)}$ is a background symmetric traceless field. 
The gauge fixing function $\Xi^{(k)}$  has been chosen such that its variation is 
\be
    \delta\Xi^{(k)}_{\mu_1\dots\mu_k} = (\square-m^2)\xi^{(k)}_{\mu_1\dots\mu_k}~.\label{General kth Gauge Fixing 2}
\ee

The partition function is given by 
\begin{equation}
    Z^{(s)} = \int\mathcal{D}(\phi;s)\Big(\prod_{k=0}^{s-1}\Delta^{(k)}\delta\left[\Xi^{(k)}-\chi^{(k)}\right]\Big)\re^{\ri S}~. \label{Spin s Partition Function 1}
\end{equation}
Since the partition function  is independent of the background fields $\chi^{(k)}$, we can average over them with a convenient weight of the form
\bea
    \prod^{s-1}_{k=0}\exp\left\{-\frac{\ri}{2}\int \rd^dx \,\Big[\frac{\chi^{(k)\mu_1\dots\mu_k}\chi^{(k)}_{\mu_1\dots\mu_k}}{\omega_k}\Big]\right\}~,
\eea
where $\omega_k$ are constants chosen such that they diagonalise the action. This gives
\begin{equation}
    Z^{(s)} =  \int\mathcal{D}(\phi;s)\left(\prod_{k=0}^{s-1}\Delta^{(k)}\exp\left\{-\frac{\ri}{2}\int \rd^dx \,\Big[\frac{\Xi^{(k)\mu_1\dots\mu_k}\Xi^{(k)}_{\mu_1\dots\mu_k}}{\omega_k}\Big]\right\}\right)\re^{\ri S}~.\label{Spin s Partition Function 2}
\end{equation}

One finds that $\Xi^{(k)\mu_1\dots\mu_k}\Xi^{(k)}_{\mu_1\dots\mu_k}$ is expanded fully as
\bea
        &&\frac{(k+1)^2(k+2)^2}{4}(\alpha_{k+1})^2\Tilde{\phi}^{(k+2)\mu_1\dots\mu_k}\Tilde{\phi}^{(k+2)}_{\mu_1\dots\mu_k}+(k+1)^2\partial_\mu\phi^{(k+1)\mu\mu_1\dots\mu_k}\partial^\nu\phi^{(k+1)}_{\nu\mu_1\dots\mu_k}\non \\
        &&+\frac{k(k+1)^2}{4}\partial^\mu\Tilde{\phi}^{(k+1)\mu_2\dots\mu_k}\partial_\mu\Tilde{\phi}^{(k+1)}_{\mu_2\dots\mu_k}+\frac{k(k+1)^2(k-1)}{4}\partial^\mu\Tilde{\phi}^{(k+1)\nu\mu_3\dots\mu_k}\partial_\nu\Tilde{\phi}^{(k+1)}_{\mu\mu_3\dots\mu_k}\non \\
        &&+\frac{(k+1)^2k^2(d+2k-4)^2}{4}(\beta_{k+1})^2\phi^{(k)\mu_1\dots\mu_k}\phi^{(k)}_{\mu_1\dots\mu_k}\non \\
        &&-\frac{(k+1)^2k^3(k-1)(d+2k-4)}{8}(\beta_{k+1})^2\Tilde{\phi}^{(k)\mu_3\dots\mu_k}\Tilde{\phi}^{(k)}_{\mu_3\dots\mu_k}\label{Square 2 Right}\\
        &&-(k+2)(k+1)^2\alpha_{k+1}\Tilde{\phi}^{(k+2)\mu_1\dots\mu_k}\partial^\mu\phi^{(k+1)}_{\mu\mu_1\dots\mu_k}+\frac{(k+2)(k+1)^2k}{2}\alpha_{k+1}\Tilde{\phi}^{(k+2)\mu_1\dots\mu_k}\partial_{\mu_1}\Tilde{\phi}^{(k+1)}_{\mu_2\dots\mu_k}\non \\
        &&+\frac{(k+2)(k+1)^2k(d+2k-4)}{2}\alpha_{k+1}\beta_{k+1}\Tilde{\phi}^{(k+2)\mu_1\dots\mu_k}\phi^{(k)}_{\mu_1\dots\mu_k}\non \\
        &&+k(k+1)^2\partial_\mu\phi^{(k+1)\mu\mu_1\dots\mu_k}\partial_{\mu_1}\Tilde{\phi}^{(k+1)}_{\mu_2\dots\mu_k}+(k+1)^2k(d+2k-4)\beta_{k+1}\partial_\mu\phi^{(k+1)\mu\mu_1\dots\mu_k}\phi^{(k)}_{\mu_1\dots\mu_k}\non \\
        &&+\frac{(k+1)^2k^2(d+2k-4)}{2}\beta_{k+1}\partial^{\mu_1}\Tilde{\phi}^{(k+1)\mu_2\dots\mu_k}\phi^{(k)}_{\mu_1\dots\mu_k}~.\non 
 \eea
These terms can be split into massless and massive parts, allowing one to look at how they modify the massive and massless Lagrangian contributions separately. Denoting the new contributions with a prime, the effect of (\ref{Square 2 Right}) on (\ref{Massless Lagrangian Contribution}) is
\bea
        \mathcal{L}'_0(\phi^{(k)})  &= & -\frac{1}{2}\partial^\mu\phi^{(k)\mu_1\dots\mu_k}\partial_\mu\phi^{(k)}_{\mu_1\dots\mu_k}+\frac{k}{2}\partial_\mu\phi^{(k)\mu\mu_2\dots\mu_k}\partial^\nu\phi^{(k)}_{\nu\mu_2\dots\mu_k}\non\\
        && +\frac{k(k-1)}{4}\partial^\mu\Tilde{\phi}^{(k)\mu_3\dots\mu_k}\partial_\mu\Tilde{\phi}^{(k)}_{\mu_3\dots\mu_k}+\frac{k(k-1)}{2}\partial_\mu\partial_\nu\phi^{(k)\mu\nu\mu_3\dots\mu_k}\Tilde{\phi}^{(k)}_{\mu_3\dots\mu_k}\non\\
        && +\frac{k(k-1)(k-2)}{8}\partial_\mu\Tilde{\phi}^{(k)\mu\mu_4\dots\mu_k}\partial^\nu\Tilde{\phi}^{(k)}_{\nu\mu_4\dots\mu_k}+\frac{(k-1)k^2}{2\omega_{k-1}}\partial_\mu\phi^{(k)\mu\nu\mu_3\dots\mu_k}\partial_\nu\Tilde{\phi}^{(k)}_{\mu_3\dots\mu_k} \non\\
        && -\frac{(k-1)k^2(k-2)}{8\omega_{k-1}}\partial^\mu\Tilde{\phi}^{(k)\nu\mu_4\dots\mu_k}\partial_\nu\Tilde{\phi}^{(k)}_{\mu\mu_4\dots\mu_k}-\frac{k^2}{2\omega_{k-1}}\partial_\mu\phi^{(k)\mu\mu_2\dots\mu_k}\partial^\nu\phi^{(k)}_{\nu\mu_2\dots\mu_k}\non\\
        && -\frac{(k-1)k^2}{8\omega_{k-1}}\partial^\mu\Tilde{\phi}^{(k)\mu_3\dots\mu_k}\partial_\mu\Tilde{\phi}^{(k)}_{\mu_3\dots\mu_k}~.\label{Massless Lagrangian Contribution Prime}
\eea
To cancel out all off diagonal and divergence contributions, one needs to choose
\begin{equation}
    \omega_k = k+1~. \label{Omega}
\end{equation}
With such a choice, (\ref{Massless Lagrangian Contribution Prime}) becomes
\bea
        \mathcal{L}'_0(\phi^{(k)})  &= & -\frac{1}{2}\partial^\mu\phi^{(k)\mu_1\dots\mu_k}\partial_\mu\phi^{(k)}_{\mu_1\dots\mu_k}+\frac{k(k-1)}{8}\partial^\mu\Tilde{\phi}^{(k)\mu_3\dots\mu_k}\partial_\mu\Tilde{\phi}^{(k)}_{\mu_3\dots\mu_k}\non\\
        && +\frac{k(k-1)}{2}\partial_\mu\partial_\nu\phi^{(k)\mu\nu\mu_3\dots\mu_k}\Tilde{\phi}^{(k)}_{\mu_3\dots\mu_k}+\frac{(k-1)k}{2}\partial_\mu\phi^{(k)\mu\nu\mu_3\dots\mu_k}\partial_\nu\Tilde{\phi}^{(k)}_{\mu_3\dots\mu_k}\non\\
        && +\frac{k(k-1)(k-2)}{8}\partial_\mu\Tilde{\phi}^{(k)\mu\mu_4\dots\mu_k}\partial^\nu\Tilde{\phi}^{(k)}_{\nu\mu_4\dots\mu_k}\label{Massless Lagrangian Contribution Prime 2}\\
        && -\frac{(k-1)k(k-2)}{8}\partial^\mu\Tilde{\phi}^{(k)\nu\mu_4\dots\mu_k}\partial_\nu\Tilde{\phi}^{(k)}_{\mu\mu_4\dots\mu_k}~,\non
  \eea
and therefore
\begin{equation}
    \begin{split}
        \int d^d x\mathcal{L}'_0(\phi^{(k)})  = & \int d^dx\left[\frac{1}{2}\phi^{(k)\mu_1\dots\mu_k}\square\phi^{(k)}_{\mu_1\dots\mu_k}-\frac{k(k-1)}{8}\Tilde{\phi}^{(k)\mu_3\dots\mu_k}\square\Tilde{\phi}^{(k)}_{\mu_3\dots\mu_k}\right]~.
    \end{split}\label{Integrated Massless Lagrangian Contribution Prime}
\end{equation}
One now must check if this value of $\o_k$ diagonalises the massive Lagrangian contributions. Indeed, by combining the massive terms in (\ref{Square 2 Right}) with (\ref{k^th Massive Lagrangian Contribution}) and (\ref{Omega}), as well as substituting (\ref{Zinoviev Coefficients}) and (\ref{Alpha k}), one has
\be
        \mathcal{L}'_m(\phi^{(k)})  =  \left[-\frac{1}{2}m^2\phi^{(k)\mu_1\dots\mu_k}\phi^{(k)}_{\mu_1\dots\mu_k}+\frac{k(k-1)}{8}m^2\Tilde{\phi}^{(k)\mu_3\dots\mu_k}\Tilde{\phi}^{(k)}_{\mu_3\dots\mu_k}\right]~.
   \ee
Thus, the entire action is diagonalised by choosing (\ref{Omega}). With this, (\ref{Spin s Partition Function 2}) becomes
\bea
        Z^{(s)} &= & \int\mathcal{D}(\phi;s)\left(\prod_{k=0}^{s-1}\Delta^{(k)}\right)\non\\
        && \times\exp\left\{\sum^{s}_{k=0}\left[\:\ri \int \rd^dx\left(\frac{1}{2}\phi^{(k)\mu_1\dots\mu_k}(\square-m^2)\phi^{(k)}_{\mu_1\dots\mu_k}\right.\right.\right.\label{Spin s Partition Function 3}\\
        && \hspace{120pt}\left.\left.\left.-\frac{k(k-1)}{8}\Tilde{\phi}^{(k)\mu_3\dots\mu_k}(\square-m^2)\Tilde{\phi}^{(k)}_{\mu_3\dots\mu_k}\right)\right]\right\}~.\non
  \eea

The Faddeev-Popov determinants  $\Delta^{(k)}$ are given by
\bea
    \Delta^{(k)} = \det(\frac{\delta \Xi^{(k)}_{\mu_1\dots\mu_k} (x)}{\delta \xi^{(k)}_{\nu_1\dots\nu_k}(x')}) = \det \Big[ \hat{\d}_{\m_1  \dots \m_k}{}^{\n_1 \dots \n_k}  (\square-m^2) \d^d(x-x')
      \Big]~,
\eea
where $\hat{\d}_{\m_1  \dots \m_k}{}^{\n_1 \dots \n_k} $ denotes the Kronecker delta on the space of symmetric traceless rank-$k$ tensors.
Making use of (\ref{Determinant of M}) gives
\begin{equation}
    \Delta^{(k)} = \int\mathcal{D}\overline{\psi}^{(k)}\mathcal{D}\psi^{(k)}\exp[-\ri \int \rd^dx \,\overline{\psi}^{(k)\mu_1\dots\mu_k}(\square-m^2)\psi^{(k)}_{\mu_1\dots\mu_k}]~. \label{k^th Ghost Contribution}
\end{equation}

The full partition function is obtained by inserting (\ref{k^th Ghost Contribution}) in (\ref{Spin s Partition Function 3}), yielding
\bea
        Z^{(s)} &= & \int\mathcal{D}(\phi,\psi;s)\exp\bigg\{\:\ri\int \rd^dx\,\bigg[
        \frac{1}{2}\phi^{(s)\mu_1\dots\mu_s}(\square-m^2)\phi^{(s)}_{\mu_1\dots\mu_s}
              \non\\
        && 
              -\frac{s(s-1)}{8}\Tilde{\phi}^{(s)\mu_3\dots\mu_s}(\square-m^2)\Tilde{\phi}^{(s)}_{\mu_3\dots\mu_s}
               \label{Spin s Partition Function 4}\\
        && +   
             \sum^{s-1}_{k=0}
             \Big(\frac{1}{2}\phi^{(k)\mu_1\dots\mu_k}(\square-m^2)\phi^{(k)}_{\mu_1\dots\mu_k}
               -\frac{k(k-1)}{8}\Tilde{\phi}^{(k)\mu_3\dots\mu_k}(\square-m^2)\Tilde{\phi}^{(k)}_{\mu_3\dots\mu_k}
              \non\\
        &&
               -\overline{\psi}^{(k)\mu_1\dots\mu_k}(\square-m^2)\psi^{(k)}_{\mu_1\dots\mu_k}\Big)\bigg]
        \bigg\}~.\non
    \eea
The structure of \eqref{Spin s Partition Function 4} is similar to the effective action for massive antisymmetric tensor field models in $d$ dimensions \cite{BKP,KT}.
It is obvious from  \eqref{Spin s Partition Function 4} that the massive propagators are remarkably simple in the Feynman-like  gauge, which we have used. An alternative quantisation scheme is to make use of the unitary gauge 
\eqref{unitary}, in which the gauge freedom is absent and the theory is described by the $d$-dimensional Singh-Hagen model \eqref{d-dimSH}. The corresponding propagators for the massive integer-spin fields in four dimensions
were derived in \cite{Singh:1981aw}.

Making use of (\ref{Spin s Partition Function 4}), one can count 
the number of degrees of freedom, $n(d,s)$, using the relation $Z^{(s)} = \det^{-n(d,s)/2} \big( \Box - m^2\big)$.
A symmetric rank-$k$ tensor in $d$ dimensions has
\bea
    \binom{d+k-1}{k}
    \label{6.15}
\eea
independent components. In the case of a traceless symmetric tensor,  this should be reduced by the number of independent components of a symmetric rank-$(k-2)$ tensor. Given a double traceless symmetric tensor, 
\eqref{6.15} should be reduced by the number of of a symmetric rank-$(k-4)$ tensor. 
The total degrees of freedom are counted as
\be
    \sum^s_{k=0}\binom{d+k-1}{k}-\sum^s_{k=4}\binom{d+k-1}{k}-2\left(\sum^{s-1}_{k=0}\binom{d+k-1}{k}-\sum^{s-1}_{k=2}\binom{d+k-1}{k}\right)~,
\ee
which, after some algebra, simplifies to $n(d,s)$ given by  eq. \eqref{3.11}.

All fields in \eqref{Spin s Partition Function 4} have one and the same kinetic operator, $(\Box -m^2)$. 
This degeneracy is characteristic of ${\mathbb M}^4$. In the case of an (anti-)de Sitter background, the structure of kinetic operators will depend on the rank of a field, and the partition function will depend on the spacetime curvature. 
This will be discussed elsewhere.
\\

\noindent
{\bf Acknowledgements:}
We are grateful to Emmanouil Raptakis for useful comments on the manuscript.
The work of SMK is supported in part by the Australian Research Council, project No. DP230101629.

\end{document}